\documentclass[useAMS,usenatbib]{mn2e}
\usepackage{graphicx}
\usepackage{amssymb}
\usepackage{rotating}

\newcommand{\apj}{ApJ}

\newcommand{\MC}{\multicolumn}
\newcommand{\kms}{km~s$^{-1}$}

\newcommand{\HI}{H{\sc i}}
\newcommand{\HII}{H{\sc ii}}
\newcommand{\sunn}{$_{\odot}$}

\title[LV dwarf Cas~I: gas metallicity, extinction and distance]
{Local Volume dwarf Cas~I: gas metallicity, extinction and distance}
\author[S.A.~Pustilnik, A.L.~Tepliakova, A.S.~Vinokurov]
{S.A.~Pustilnik,$^{1}$\thanks{E-mail: sap@sao.ru (SAP)}
A.L.~Tepliakova,$^{1}$ A.S.~Vinokurov$^{1}$ \\
$^1$ Special Astrophysical Observatory of RAS, Nizhnij Arkhyz,
Karachai-Circassia 369167, Russia\\
}

\begin{document}

\label{firstpage}

\date{Accepted ??, 2023. Received December 6, 2023}

\pagerange{\pageref{firstpage}--\pageref{lastpage}} \pubyear{2023}

\maketitle

\begin{abstract} 
Cas~I is a Local Volume (LV) dwarf irregular galaxy with a wide range of suggested distances.
Tikhonov (2019), using the HST images and the Tip of Red Giant Branch (TRGB) method,
places Cas~I at $D$ = 1.6$\pm$0.1~Mpc. Besides, he estimates the stellar metallicity of Cas~I
at the level of z $\sim$ 0.0004 (or Z(stars) $\sim$ Z\sunn/50).
Such a nearby extremely low-metallicity dwarf, if real, would be a very valuable object for detailed studies.
An alternative TRGB distance of Cas~I, of 4.5$\pm$0.2~Mpc, based on the deeper HST images,
was presented in the Extragalactic Distance Database (EDD). It places Cas~I midway between
groups of IC342 (D $\sim$ 3.3~Mpc) and Maffei1 and 2 (D $\sim$ 5.7~Mpc).      %}
%{
We wish to check the suggested extremely low metallicity of Cas~I, to improve the estimate of the large
Milky Way (MW) extinction in its direction and to use this data to improve the distance estimate to Cas~I.  %}
%{
We use the SAO 6-m telescope (BTA) spectroscopy to estimate gas metallicity in two representative
\HII\ regions in Cas~I and to derive, via their observed Balmer decrements, the independent upper limit to
the value of the MW extinction.  %}
%{
We derive the values of 12+log(O/H) = 7.83$\pm$0.1~dex and 7.58$\pm$0.1~dex in two \HII\ regions of Cas~I, which
correspond to Z(gas) of $\sim$5--10~times higher than Z(stars) estimated for its resolved stars. The measured
Balmer decrements in these \HII\ regions, result in the maximal MW extinction in this direction of
A$_{\rm B}$ = 3.06$\pm$0.14~mag in comparison to A$_{\rm B}$ = 3.69$\pm$0.4 derived via the IR dust
emission and is used in other estimates of the distance to Cas~I. This reduces the original EDD distance till 4.1~Mpc.
The relation of Berg et al. (2012) for the LV late-type galaxies, between their 12+log(O/H) and blue absolute magnitude,
M$_{\rm B}$, is used to bracket M$_{\rm B}$ for Cas~I. This, in turn, allows one to get an independent estimate of
the distance to Cas~I, of $\sim$1.64~Mpc, albeit with the large 1$\sigma$ uncertainty of factor 2.17.    %}
%{
The combination of the above distance estimates, accounting for their uncertainties, results in the most probable
value of D$\sim$3.65~Mpc, what favours  Cas~I to reside in the environs of IC342.   %}
\end{abstract}


\section[]{Introduction}
\label{sec:intro}
\setcounter{figure}{0}

The galaxy population of the Local Volume (LV) is one of the best studied thanks to its proximity
and due to its importance as the benchmark sample for many extragalactic and near-field cosmology studies.
We are especilly interested in the better understanding of the properties of the LV galaxies
which reside in the Nearby Voids, that is are a subgroup of the Nearby Void Galaxy (NVG)
sample from \citet{PTM19}. The recent results of our probing the gas metallicity
of the NVG sample galaxies are presented in papers of \citet{XMP-SALT, XMP-BTA} and
in Pustilnik et al (2023, MNRAS, in press). Eleven of these galaxies were found to be extremely 
ow-metallicity, with 12+log(O/H) $\sim$ 7.0 -- 7.2~dex, or Z(gas) $\sim$ Z\sunn/50 -
- Z\sunn/30.\footnote{We adopt the solar Oxygen abundance of 12+log(O/H) = 8.69~dex as in \citet{Asplund2009}}

The LV dIrr galaxy Cas~I, since its discovery in the Zone of Avoidence (ZoA)
by \citet{Weinberger95}, has a long story of its distance determination.
The suggested estimates of the distance range from 0.8 to 4.5~Mpc
\citep{Tikhonov96,Karachentsev97,WS1998,UNGC,Tikhonov19,Anand21}.

In the time, when we collected  the NVG sample, Cas~I was assigned to
belong to the environs of the luminous spiral IC342 \citep{UNGC}, situated at the distance
of 3.3$\pm$0.13~Mpc \citep{Anand21}, and therefore did not enter to the NVG sample.
After the NVG sample was published, \citet{Tikhonov19} presented an alternative distance of Cas~I.
Based on the Hubble Space Telescope (HST) Colour-Magnitude Diagram (CMD), he used the
well-known Tip of Red Giant branch (TRGB) method to derive its distance of D=1.6$\pm$0.1~Mpc.
While this distance appears substantially smaller than that of IC342 group, at this position
Cas~I falls to a rather isolated environment within the boundaries of the void~No.8 (Ori-Tau)
from the list of the nearby voids \citep{PTM19}.

Besides, via the fitting of the CMD with the theoretical stellar isochorones,
\citet{Tikhonov19} finds the very low stellar metallicity in this galaxy, of z = 0.0004 -- 0.0007
(or Z $\sim$Z\sunn/50 -- Z\sunn/30).

Our primary motivation was to clear up the issue of Cas~I very low metallicity as possibly related
to its assumed the low-density environment. Besides, Cas~I is situated at the low galactic
latitude of b$^{II}$ = +7.1\degr\ (in the ZoA), where estimates of the MW extinction have rather large
uncertainties. The commonly accepted MW extinction is based on the paper by \citet{Schlafly11}.
This gives for Cas~I the value of E$(B-V)$ = 0.900~mag.
The related values of A$_{\rm B}$ =  4.1$\times$E$(B-V) \sim$3.7~mag and A$_{\rm V} \sim$2.8~mag,
have, according to  \citet{Schlegel98}, the internal uncertainties of at least 10 per cent. That is,
for the case of Cas~I, $\sigma$(A$_{\rm B}$,A$_{\rm V}$) $\gtrsim$ 0.3--0.4~mag.

This rather large uncertainty may affect the determination of Cas~I global parameters and its distance.
The only independent estimate of A$_{\rm V} \sim$2.5~mag \citep{WS1998} is based on the Balmer decrement in
one of Cas~I \HII\ regions, derived from the observed flux ratio of only two lines, H$\alpha$ and H$\beta$.
Therefore, the second task of this work was to obtain an independent upper limit of the MW extinction
via the new measurement of Balmer decrement in two different \HII\ regions.

Meantime, the results of the recent HST imaging, about 1.5~mag deeper than the data used by \citet{Tikhonov19},
were presented in the Extragalactic Distance Database (EDD)\footnote{https://\mbox{edd.ifa.hawaii.edu/dfirst.php?}}.
They give an alternative TRGB distance of 4.5$\pm$0.2~Mpc. Their estimate seems does not take
into account the error of the adopted extinction in $I$ band, minimum of 0.15~mag, according to \citet{Schlegel98}.

The spectra of \HII\ regions allow one to obtain an independent estimate of the sight-line extinction via the observed
Balmer decrement, and thus, to potentially improve the accuracy of the derived absolute magnitude and other affected
parameters of Cas~I.
Therefore, the attempt to derive an independent estimate
of the MW extinction with as high as possible accuracy appears rather actual.
To this end, we obtained spectra of two  \HII\ regions, seen in the available images of its H$\alpha$ emission
\citep{Kaisin2019} and also marked by \citet{Tikhonov19} on the HST images.

The rest of the paper is arranged as follows. In Sec.~\ref{sec:obs}, the spectral observations and data reduction
are outlined. In Sec.~\ref{sec:results}, the results of the analysis of the BTA spectra are presented.
In Sec.~\ref{sec:discuss}, we discuss properties of the studied star-forming regions in Cas~I, including their
gas metallicity and extinction as well as the application of the reference relation of  O/H(gas) versus M$_{\rm B}$
from \citet{Berg12}. In Sec.~\ref{sec:summary} we summarize our results and conclude.
The linear scale at the {\it adopted} distance to Cas~I of 3.6~Mpc is
$\sim$17.5~pc in 1~arcsec.

\section[]{Observations and data processing}
\label{sec:obs}

\begin{table}
\centering{
\caption{Journal of BTA observations of Cas~I}
\label{tab:journal}
\hoffset=-2cm
\begin{tabular}{l|l|l|c|c} \hline  \hline \\ [-0.2cm]
\MC{1}{c|}{Date} &
\MC{1}{c|}{Grism}&
\MC{1}{c|}{Expos.}&
\MC{1}{c|}{$\beta$}&
\MC{1}{c}{Air}  \\

\MC{1}{c|}{ } &
\MC{1}{c|}{} &
\MC{1}{c|}{time, s}&
\MC{1}{c|}{arcsec} &
\MC{1}{c}{mass}\\

\hline     % \\
2023.10.22 & VPHG1200B  & 4$\times$800   & 1.7 & 1.10 \\ % PA=17.2; mean PA parallactic = -7.0
2023.10.22 & VPHG1200R  & 2$\times$600   & 1.7 & 1.11 \\ % PA=17.2
\hline \hline \\ %[-0.2cm]
\end{tabular}
}
\end{table}

\begin{figure}
\centering{
\includegraphics[width=7.5cm,angle=-0,clip=]{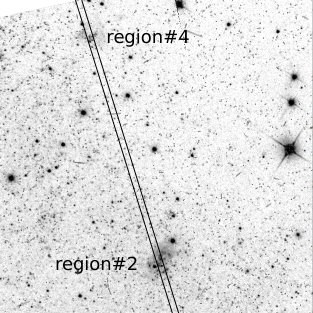}
\caption{SCORPIO-1 slit position (PA = 17.2\degr) overlaid on the
HST F606W image. Two target \HII-regions ($\#$2 and $\#$4) are marked according to
the nomenclature from \citet{Tikhonov19}. The side of the image is 1~arcmin.
 } }
\label{fig:BTA-slit}
\end{figure}

We obtained two optical spectra of Cas~I with the BTA multimode
instrument SCORPIO-1 \citep{SCORPIO} during the night 2023 October 22,
under photometric conditions and seeing of 1.7~arcsec (see Table~\ref{tab:journal}).
The long slit with the length of 6~arcmin and width of 1.2~arcsec, with the scale along
the slit of $\sim$0.36~arcsec pixel$^{-1}$ (after binning by 2) was positioned on the
Cas~I \HII\ region $\#$2 (in the nomenclature of  \citet{Tikhonov19}).
% The resulting spectral resolution was of FWHM $\sim$ 6.0~\AA.

To minimize the light loss due to the differential atmospheric refraction \citep[e.g.][]{Filippenko82},
the position andgle (PA) of the slit is recommended to keep close to the direction of atmosperic refraction,
that is close to the parallactic position angle, PA$_{\rm par}$.
We observed Cas~I near the meridian, so that the average PA$_{\rm par}$ during observations was \mbox{$\sim$--7\degr}.
The air masses during the time of this observation was rather small: of $\sim$1.10--1.11. Therefore,
the relatively small difference of the real PA and that of PA$_{\rm par}$ should give the negligible
effect on the resulting relative line fluxes.

\begin{figure}
\centering{
\includegraphics[width=6.0cm,angle=-90,clip=]{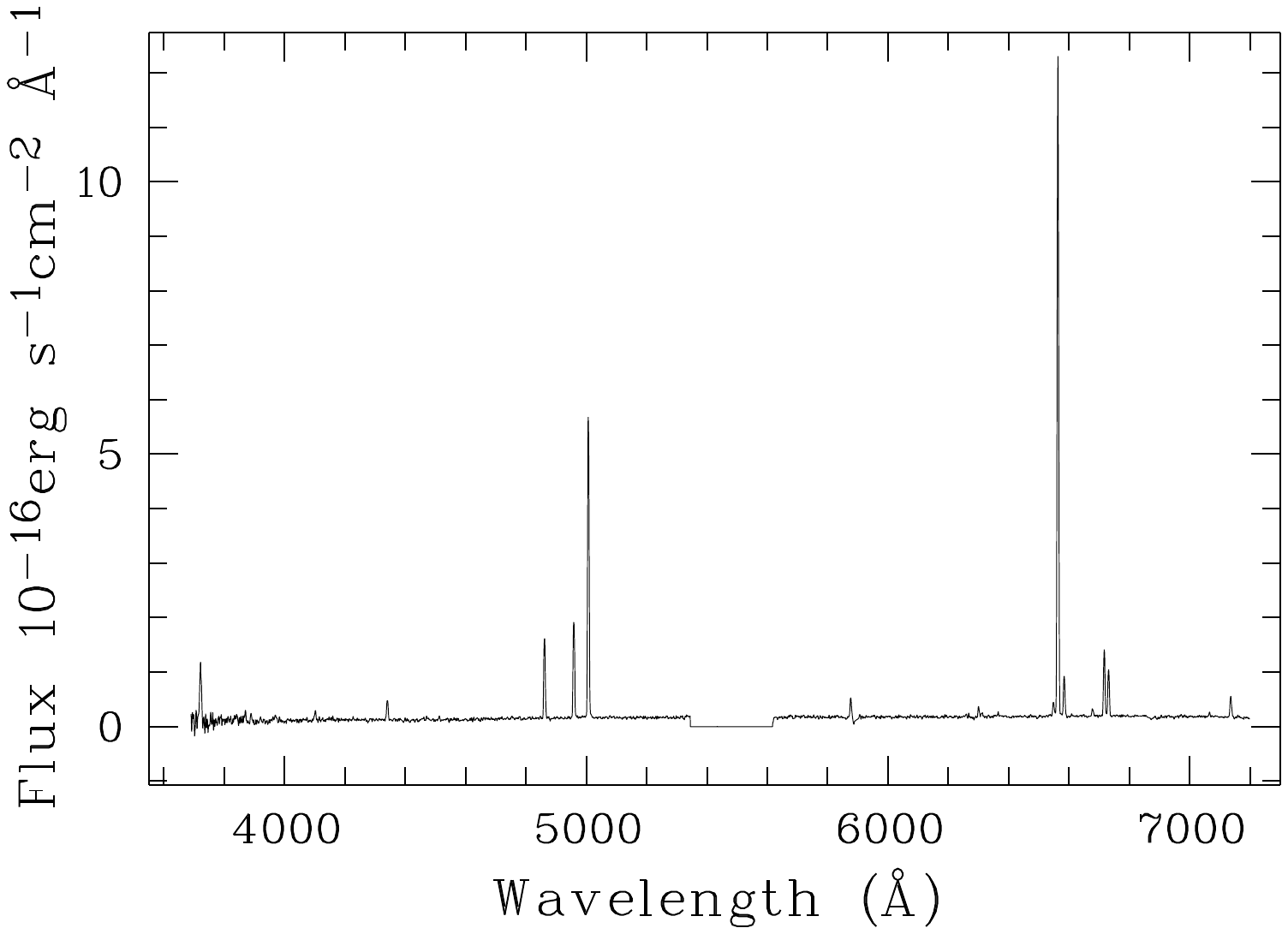}
\includegraphics[width=6.0cm,angle=-90,clip=]{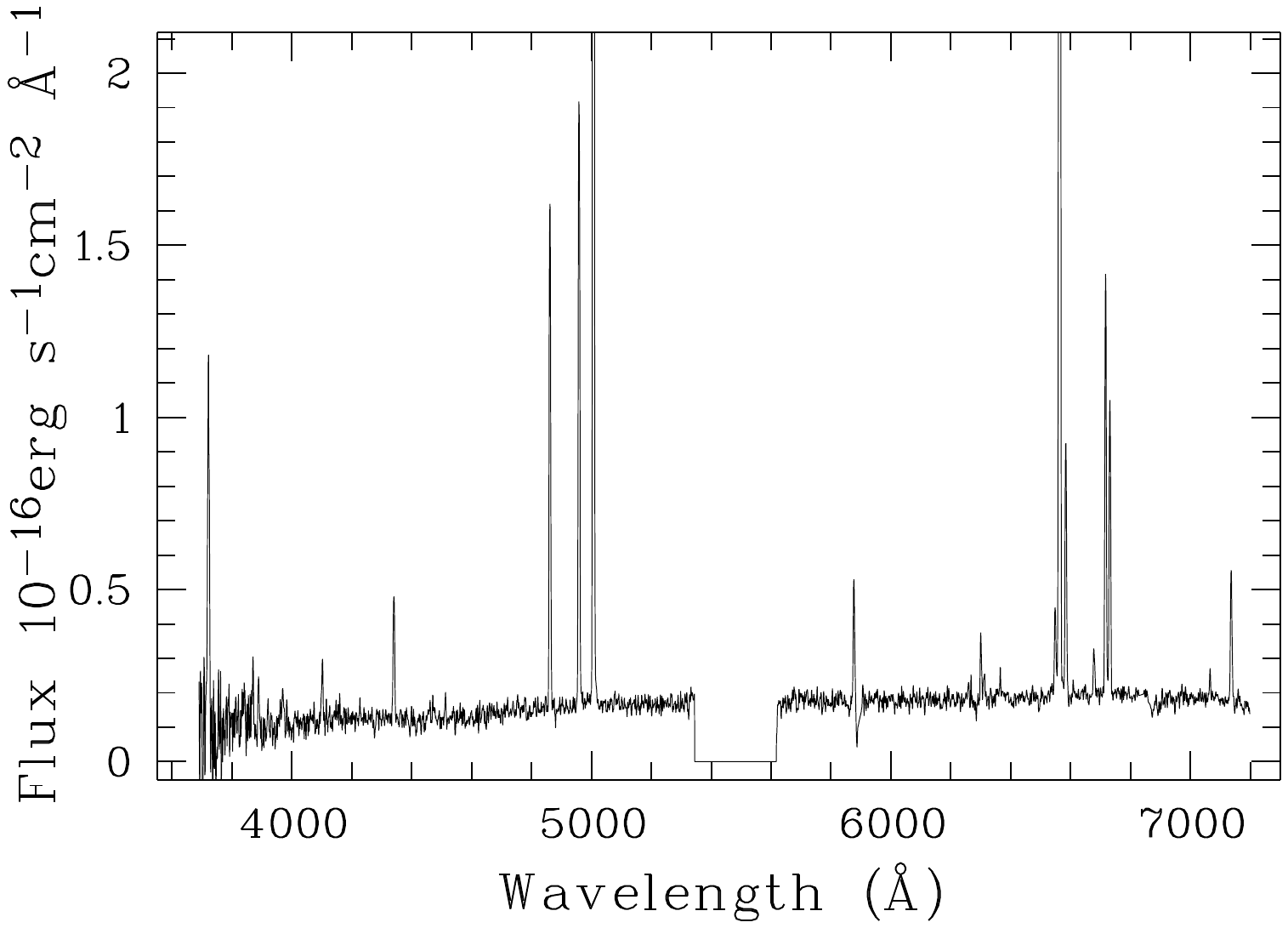}
\caption{BTA spectrum of 2023.10.22 of \HII\ region $\#$2 in galaxy Cas~I.
{\bf Top  panel:} The whole 1D spectrum.
{\bf Bottom panel:} The close-up of the same 1D spectrum to show weak features.
 } }
\label{fig:Spectra_N2}
\end{figure}

\begin{figure}
\centering{
\includegraphics[width=6.0cm,angle=-90,clip=]{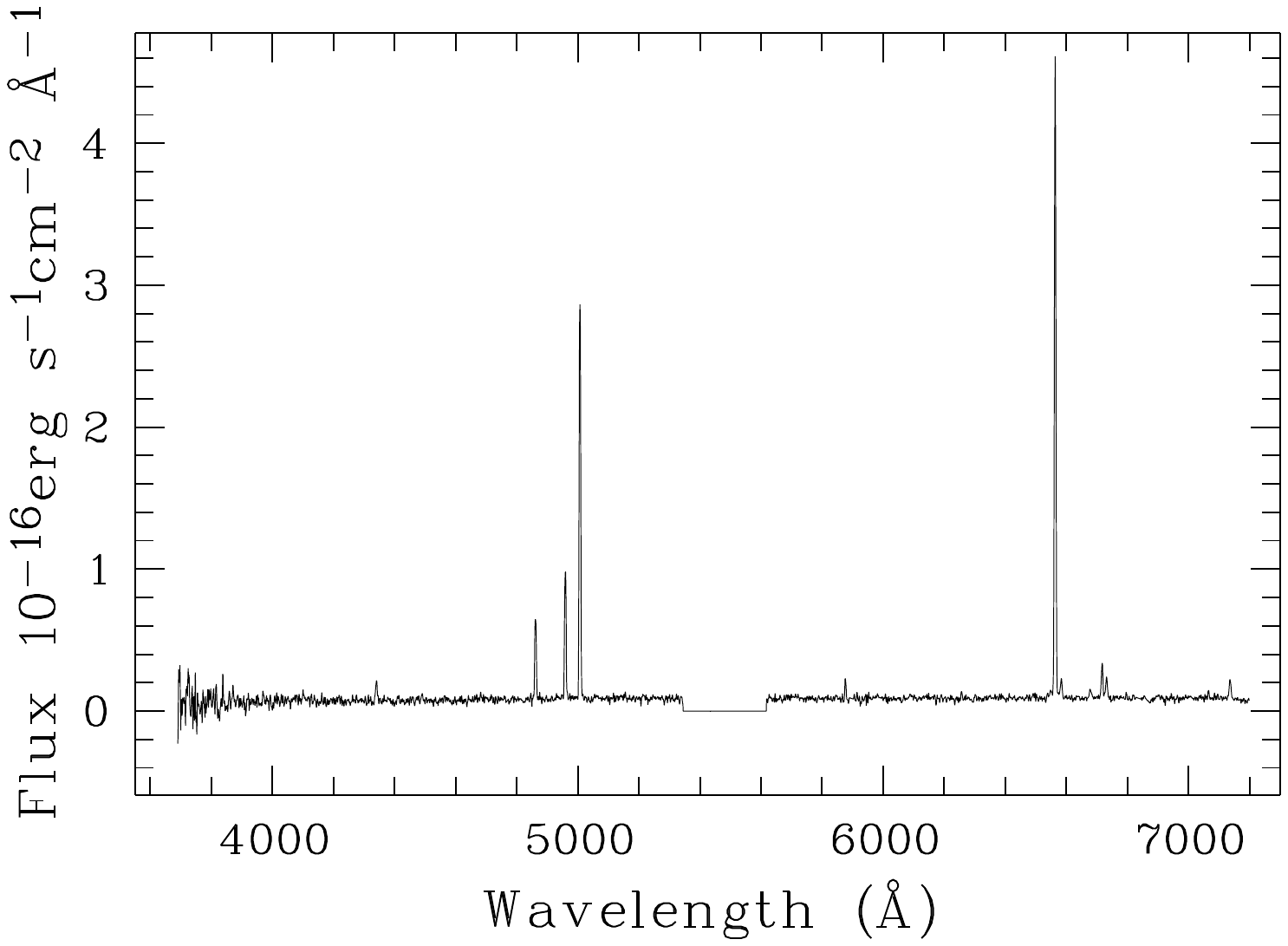}
\includegraphics[width=6.0cm,angle=-90,clip=]{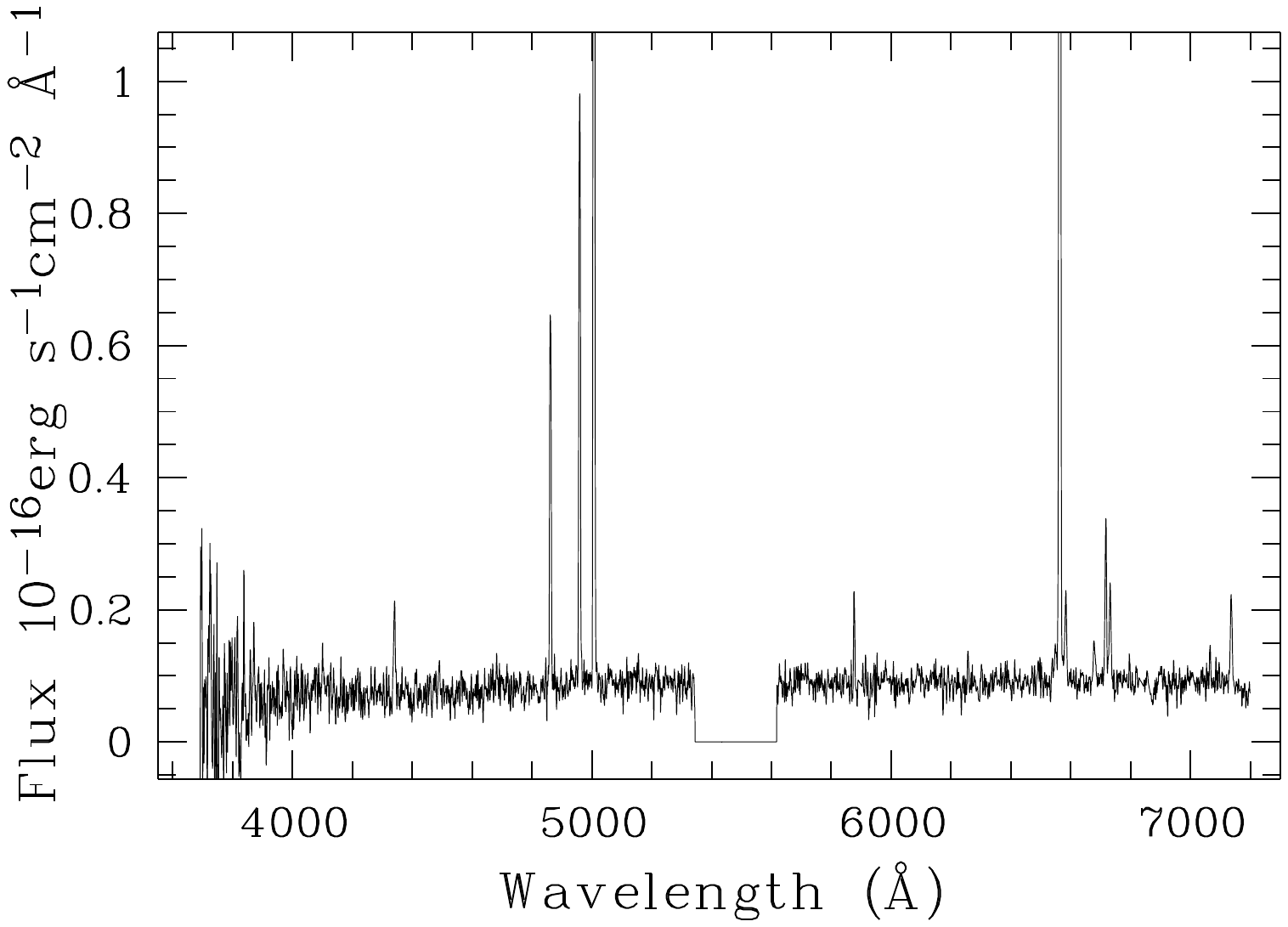}
\caption{BTA spectrum of 2023.10.22 of \HII\ region $\#$4 in galaxy Cas~I.
{\bf Top  panel:} The whole 1D spectrum.
{\bf Bottom panel:} The close-up of the same 1D spectrum to show weak features.
 } }
\label{fig:Spectra_N4}
\end{figure}

With these estimates in hand, we put the slit PA = 17.2\degr\ so that the slit crossed another relatively bright
\HII\ region of Cas~I, namely $\#$4 at $\sim$50~arcsec to the NE.
See Fig.~\ref{fig:BTA-slit} with the slit position superimposed on the HST F606W image, taken from the
EDD data base. The grisms VPHG1200B and VPHG1200R with the 2K$\times$2K CCD detector E2V~42-40
(13.5$\times$13.5~$\mu$m pixel) provided the spectrum coverage of $\sim$3650--5450~\AA\ and 5700--7500~\AA, respectively,
with the spectral resolution of FWHM $\sim$5.5~\AA. We obtained spectra with both grisms at the same slit
position and under the similar conditions.

Due to a bit different value of seeing, one expects the related different light loss on the slit for blue and red grisms.
Indeed, we found that the continuum level in the red part of spectrum runs a bit higher than its continuation from
the 'yellow' edge.
To eliminate, at first approximation, this difference between the two parts of the spectrum, before to combine both
parts, we multiply the red part fluxes by a factor of 0.89.

The main procedures of the data reduction are described in \citet{PaperVII}. Here we briefly outline them.
Our standard pipeline with the use of {\tt IRAF}\footnote{IRAF: the Image Reduction and Analysis Facility
is distributed by the National Optical Astronomy Observatory, which is operated by the Association of Universities
for Research in Astronomy, Inc. (AURA) under cooperative agreement with the National Science Foundation (NSF).}
and {\tt MIDAS}\footnote{MIDAS is an acronym for the European Southern Observatory package -- Munich Image Data
Analysis System. }
was applied for reduction of the long-slit spectra. It includes the following steps: removal of cosmic ray hits,
bias subtraction, flat-field correction, wavelength calibration, night-sky background subtraction.
Spectrophotometric standard star BD+28\degr4211, observed before Cas~I, was used for the flux calibration.

In the resulting 2d spectra with both grisms, the main bright emission lines in both \HII\ regions,
$\#$2 and $\#$4, are  well seen.
The 1d spectra of these regions were extracted, summing up 21 pixels ($\sim$7.5~arcsec) and 19 pixels ($\sim$6.8~arcsec),
respectively, without weights, centred on the maximum of the H$\gamma$ line signal.

The resulting 1d spectra are shown in the top and bottom (close-up to
show the level of continuum and the faint emission lines) panels of
Fig.~\ref{fig:Spectra_N2} for region $\#$2 and Fig.~3  % \ref{fig:Spectra_N4}
-- for $\#$4.

\begin{table*}
\centering{
\caption{Observed and derived parameters of  Cas~I \HII\ regions \#2 and \#4}
\label{t:Intens_line}
\begin{tabular}{lcccc} \hline
\rule{0pt}{10pt}
Region                    & \MC {2}{c}{$\#$2}                    &  \MC {2}{c}{$\#$4}               \\
\hline
$\lambda_{0}$(\AA) Ion    & F($\lambda$)/F(H$\beta$)&I($\lambda$)/I(H$\beta$) & F($\lambda$)/F(H$\beta$)&I($\lambda$)/I(H$\beta$)                          \\
\hline
3727\ [O\ {\sc ii}]\      & 0.868$\pm$0.055 & 1.994$\pm$0.142    & 0.317$\pm$0.086 & 0.670$\pm$0.205  \\
4101\ H$\delta$\          & 0.101$\pm$0.010 & 0.295$\pm$0.051    & 0.108$\pm$0.019 & 0.304$\pm$0.070  \\
4340\ H$\gamma$\          & 0.268$\pm$0.036 & 0.457$\pm$0.071    & 0.258$\pm$0.023 & 0.447$\pm$0.049  \\
4861\ H$\beta$\           & 1.000$\pm$0.033 & 1.000$\pm$0.038    & 1.000$\pm$0.042 & 1.000$\pm$0.049  \\
4959\ [O\ {\sc iii}]\     & 1.222$\pm$0.038 & 1.048$\pm$0.037    & 1.548$\pm$0.059 & 1.322$\pm$0.057  \\
5007\ [O\ {\sc iii}]\     & 3.831$\pm$0.203 & 3.179$\pm$0.100    & 4.739$\pm$0.164 & 3.927$\pm$0.153  \\
5876\ He {\sc i}   \      & 0.222$\pm$0.009 & 0.108$\pm$0.005    & 0.188$\pm$0.017 & 0.095$\pm$0.010 \\
6300\ [O\ {\sc i}]  \     & 0.094$\pm$0.007 & 0.037$\pm$0.003    & 0.043$\pm$0.018 & 0.017$\pm$0.007 \\
6312\ [S{\sc iii}] \      & 0.042$\pm$0.006 & 0.016$\pm$0.002    & ...             & ...             \\
6548\ [N\ {\sc ii}]\      & 0.170$\pm$0.029 & 0.058$\pm$0.011    & 0.083$\pm$0.065 & 0.031$\pm$0.027 \\
6563\ H$\alpha$\          & 8.189$\pm$0.200 & 2.827$\pm$0.084    & 7.508$\pm$0.244 & 2.788$\pm$0.111 \\
6584\ [N\ {\sc ii}]\      & 0.512$\pm$0.034 & 0.173$\pm$0.013    & 0.261$\pm$0.073 & 0.095$\pm$0.030 \\
6678\ He {\sc i}   \      & 0.089$\pm$0.006 & 0.029$\pm$0.002    & 0.140$\pm$0.019 & 0.049$\pm$0.008 \\
6717\ [S\ {\sc ii}]\      & 0.819$\pm$0.027 & 0.267$\pm$0.009    & 0.386$\pm$0.020 & 0.135$\pm$0.007 \\
6731\ [S\ {\sc ii}]\      & 0.563$\pm$0.018 & 0.183$\pm$0.006    & 0.267$\pm$0.019 & 0.093$\pm$0.007 \\
7136\ [Ar\ {\sc iii}]\    & 0.270$\pm$0.010 & 0.072$\pm$0.003    & 0.250$\pm$0.018 & 0.072$\pm$0.006 \\
& & \\
C(H$\beta$)\ dex          & \MC {2}{c}{1.295$\pm$0.032}          & \MC {2}{c}{1.190$\pm$0.042}    \\
EW(abs)\ \AA\             & \MC {2}{c}{4.90$\pm$0.60}            & \MC {2}{c}{4.55$\pm$0.63}    \\
F(H$\beta$)$^a$\          & \MC {2}{c}{8.99$\pm$0.19}            & \MC {2}{c}{3.72$\pm$0.11}      \\
EW(H$\beta$)\ \AA\        & \MC {2}{c}{54.85$\pm$1.30}           & \MC {2}{c}{45.51$\pm$1.40}     \\
 Rad. vel.\ \kms\         & \MC {2}{c}{  20$\pm$12}              & \MC {2}{c}{  28$\pm$6 }        \\ \hline % after zero-point correct. of ~ -0.25~A  (-11.5 km/s) and heliocent. +1.4 km/s
& \\
$T_{\rm e}$(OIII)(K)\                & \MC {2}{c}{14411$\pm$1169~~} & \MC {2}{c}{17230$\pm$1339~~}  \\
$T_{\rm e}$(OII)(K)\                 & \MC {2}{c}{13584$\pm$678~~}  & \MC {2}{c}{14734$\pm$316~~}   \\
$T_{\rm e}$(SIII)(K)\                & \MC {2}{c}{14018$\pm$1252~~} & \MC {2}{c}{16604$\pm$1022~~}  \\
$N_{\rm e}$(SII)(cm$^{-3}$)\         & \MC {2}{c}{10$\pm$10}         & \MC {2}{c}{10$\pm$10}          \\
O$^{+}$/H$^{+}$($\times$10$^5$)\     & \MC {2}{c}{2.487$\pm$0.452~~} & \MC {2}{c}{0.643$\pm$0.210~~}  \\
O$^{++}$/H$^{+}$($\times$10$^5$)\    & \MC {2}{c}{3.861$\pm$0.818~~} & \MC {2}{c}{3.105$\pm$0.558~~}  \\
O/H($\times$10$^5$)\                 & \MC {2}{c}{6.348$\pm$0.935~~} & \MC {2}{c}{3.748$\pm$0.596~~}  \\
12+log(O/H)(mse,c)\                  & \MC {2}{c}{~7.81$\pm$0.12~~}  & \MC {2}{c}{~7.59$\pm$0.12~~}   \\
12+log(O/H)(PT05)\                   & \MC {2}{c}{~7.85$\pm$0.10~~}  & \MC {2}{c}{~7.56$\pm$0.10~~}   \\
\hline
%\MC{3}{l}{~~} \\
\MC{3}{l}{$^a$ in units of 10$^{-16}$ ergs\ s$^{-1}$cm$^{-2}$.}\\
\rule{0pt}{10pt}
\end{tabular}
}
\end{table*}

\section{Results}
\label{sec:results}

\subsection{Emission line parameters}
\label{ssec:emis.lines}

The results of emission line measurements and analysis for the
both spectra are presented in Table~\ref{t:Intens_line}.
In the left column we give the measured line fluxes relative to that of H$\beta$.
In the right column we give the relative line intensities corrected for
the extinction
% (which is the sum of the Milky Way and that of the internal extinction
% in in the observed  \HII\ regions of Cas~I)
and the underlying Balmer line absorbtion in the stellar continuum of the both
\HII\ regions. The latter procedure is performed by the method described
in \citet{Izotov94}, via variation of both C(H$\beta$) and EW(abs) in the
wide ranges and finding the best combination of the two parameters for a given
electron temperature T$_{\rm e}$. Then, with several iterations of T$_{\rm e}$,
the best combination T$_{\rm e}$, C(H$\beta$) and EW(abs) is found that reproduces
the theoretical intensities of Balmer emissions in Case~B recombination, in terms
of minimum of $\chi^2$.

We recall that in further we assume the two-zone model of an \HII\ region as described
by  \citet[e.g.][]{Stasinska90, Izotov94}.

Since the faint temperature-sensitive auroral line [O{\sc iii}]$\lambda$4363 is not
detected in our spectra, we can not use the direct method of determination of T$_{\rm e}$.
Therefore, we employ the modified (see below) semi-empirical method of \citet{IT07} and
the empirical method \citet{PT05} to derive the parameter 12+log(O/H) from the available
line intensities.

The semi-empirical method of \citet{IT07} is based on the emprical relation between
T$_{\rm e}$ and the parameter R$_{\mathrm {23}}$. The latter is the ratio of
[I([O{\sc ii}]3727) + I([O{\sc iii}]4959) + I([O{\sc iii}]5007)]/I(H$\beta$).
In this method, on the first step the parameter T$_{\rm e}$ is estimated via parameter
R$_{\mathrm {23}}$, and then O/H is calculated with this T$_{\rm e}$ as in the classical
direct method \citep[e.g.][]{Izotov06}.

We revised the method of \citet{IT07} in \citet{XMP-BTA} to account for the substantial dependence of the derived
T$_{\rm e}$ on the excitation parameter O$_{\mathrm {32}}$. The latter is defined as the ratio of line fluxes
I([O{\sc iii}]5007)/I([O{\sc ii}]3727). This modified semi-empirical method (called hereafter 'mse') is applicable
in the range of 12+log(O/H) $\sim$ 7.0--8.1~dex. % \citep{XMP-BTA}.
Its internal scatter $\sigma$(log(O/H),mse) = 0.09~dex.
% {\bf Add about empirical relation of T$_{\rm e}$(O{\sc iii}) and T$_{\rm e}$(O{\sc ii}).}

The second method is the well known empirical method of \citet{PT05}. It uses both
parameters R$_{\mathrm {23}}$ and P, the analog of the above excitation parameter O$_{\mathrm {32}}$.
Their parameter P is defined as P = R$_{\mathrm {3}}$/R$_{\mathrm {23}}$, where
R$_{\mathrm {3}}$ = [I([O{\sc iii}]4959)+I([O{\sc iii}]5007)]/I(H$\beta$).
In terms of O$_{\mathrm {32}}$, we can write P = O$_{\mathrm {32}}$/(0.75+O$_{\mathrm {32}}$).
%  Or inverting this, we get the relation O32 = 0.75*P/(1-P). That is PT05 is applicable for the range of
%  O32 = [0.917 - 24.25]
%The latter relation accounts for the theoretical ratio of I([O{\sc iii}]5007)/I([O{\sc iii}]4959) = 2.98 {\bf (REF)}.
The internal scatter of the \citet{PT05} method is $\sim$0.1~dex.
We use their formula (24) for the so-called lower branch,  what is valid for the range
of 12+log(O/H) $\lesssim$ 8.2~dex. The method is well calibrated for the range of P = 0.55--0.97.
Our case, with P $\sim$ 0.68 and 0.89, and 12+log(O/H) $\sim$ 7.6 --  7.8~dex, corresponds well to these limitations.
%% O32(a) = 1.594 P(a) = 0.68
%% O32(b) = 5.861 P(b) = 0.89

One of the by-products of the \HII\ spectrum analysis is the independent
determination of the extinction inherent to the studied regions.
The derived values of C(H$\beta$) are 1.295$\pm$0.032 and 1.190$\pm$0.042, for regions $\#$2 and $\#$4, respectively.
Since they are close within their uncertainties, we adopt the average C(H$\beta$)$_{\rm mean}$ = 1.25$\pm$0.05.
Then, using the well-known relation between C(H$\beta$) and the colour excess:  $E(B-V)$ = 0.68 $\times$ C(H$\beta$),
%(NEED REF),
we obtain $E(B-V)$ = 0.85 $\pm$ 0.034.
Then, from the relation of A$_{\rm B}$ = 4.1~$E(B-V)$, we obtain
A$_{\rm B}$ = 3.48 $\pm$ 0.14~mag and A$_{\rm V}$ = 2.64 $\pm$ 0.11~mag.
The latter value of A$_{\rm B}$ is somewhat smaller than A$_{\rm B, MW}$ = 3.69$\pm$0.37~mag, related to the Milky Way
extinction in this direction according to \citet{Schlafly11}. However, the two values are consistent within the cited
uncertainties.

It is worth  mentioning the value of extinction in \HII\ region $\#$2, derived by \citet{WS1998}.
From the ratio of line fluxes of H$\alpha$ and H$\beta$, they derive the value of A$_{\rm V}$ = 2.5.
Judging from the S-to-N in the H$\beta$ line, the probable error of their A$_{\rm V}$ is at least 5 per cent, or 0.12~mag.
Their A$_{\rm V}$ is somewhat smaller than our, but is consistent with our A$_{\rm V}$ = 2.64~mag within one combined error
of $\sim$0.15~mag.

Thus, summarising, our estimate of extinction in the direction of Cas~I, based on the Balmer decrement
in two \HII\ regions, is midway between the two independent estimates
via the dust emission \citep{Schlafly11} and via the Balmer decrement in the same \HII\ region $\#$2 \citep{WS1998}.
Having in mind that our estimate of A$_{\rm B}$ has the error of at least 2.5 times smaller than A$_{\rm B}$,
predicted in paper by \citet{Schlafly11}, we will use in further discussion A$_{\rm B}$ = 3.48 $\pm$ 0.14~mag and
the related extinction in other bands.

%The extinction presented in NED, A$_{\rm B}$ = 3.69, is derived by \citet{Schlafly11}, based on
%the maps of IR emission of the MW dust.
%The galactic lattitude of Cas~I of $b^{II}$ = +7.1\degr\ is quite
%close to the boundary of $|b^{II}| < $ 5\degr, below which the minimal uncertainty of A$_{\rm B}$ is 10\%.
%Therefore, we adopt the probable error of their estimate as of $\gtrsim$0.3~mag.
%Thus, within the cited uncertainties, our spectroscopic estimate is well consistent with that
%from \citet{Schlafly11} based on the IR emission of the MW dust.

\begin{table}
\caption{Properties of Cas~I}
\begin{tabular}{lcc}
\hline
Property                             & Value         & Refs          \\
\hline
RA (J2000)                           & 02 06 05.39   & 2            \\
Dec (J2000)                          & +62 00 12     & 2             \\
Rad. velocity ($\#$2,$\#$4), \kms\   & 32$\pm$5      & 1             \\
Rad. velocity (\HI), \kms\           & 35$\pm$1      & 2             \\
$D_{\rm adopt(TRGB+O/H)}$, Mpc       & 3.65           & 1             \\
$D$(Tikhonov 2019), Mpc              & 1.6$\pm$0.1   & 3             \\
$D$(EDD, original), Mpc              & 4.51$\pm$0.22  & 4          \\
$D$(EDD, corrected), Mpc             & 4.11$\pm$0.36   & 1        \\
B$_{\rm tot}^1$ (mag)                & 15.30         & 2             \\
A$_{\rm B}$ (mag)                    & 3.69$\pm$0.37 & 5             \\
A$_{\rm B}$(Balmer) (mag)            & 3.48$\pm$0.14 & 1             \\
A$_{\rm B}$(Balmer, corrected) (mag) & 3.20$\pm$0.14 & 1             \\
12+log(O/H)($\#$2)                  & 7.83$\pm$0.10 & 1             \\
12+log(O/H)($\#$4)                  & 7.59$\pm$0.10 & 1             \\
12+log(O/H)(aver.$\#$2,$\#$4)        & 7.71$\pm$0.10 & 1             \\
$M_{\rm B}$(mag)(12+log(O/H)=7.71) &--13.44$\pm$1.68& 1       \\
$\mu_{\rm 0,B}$ (mag~arcsec$^2$)     & 23.3          & 2             \\
S(\HI) (Jy~\kms)                     & 50.0          & 2             \\
M(\HI) (in 10$^7$ M\sunn)            & 16.1(D/3.7)$^2$& 1             \\
M(\HI)/L$_{\rm B}$ (M\sunn/L\sunn)   & 0.53          & 1             \\
% M(*) (in 10$^6$ M\sunn)            & ??.0          & ?             \\
\hline
\multicolumn{3}{p{7.0cm}}{
{\bf 1.} This work
{\bf 2.} \citet{UNGC} and its update: \mbox{https://www.sao.ru/lv/lvgdb}.
{\bf 3.} \citet{Tikhonov19}
{\bf 4.} \mbox{http://edd.ifa.hawaii.edu}
{\bf 5.} The Milky Way extinction after \citet{Schlafly11}.
}
\end{tabular}
\label{tab:parameters}
\end{table}

\section{Discussion}
\label{sec:discuss}

\subsection{Environment}
\label{ssec:environs}

The dwarf galaxy Cas~I resides rather close in projection to two massive galaxies,
IC342  at the angular distance of $\sim$5\degr\ and Maffei1 (PGC009892) at $\sim$3.5\degr.
The differences of their heliocentric                  % V(\HI) with the heliocentric
velocities are also rather small, of 48~\kms\ and 31~\kms, respectively.
Groups of IC342 and Maffei1/Maffei2 are currently considered as the different aggregates
with the distance of IC342 of 3.3~Mpc \citep{Anand21} and the distance of Maffei1/Maffei2
of 5.73~Mpc. The latter distance was first revised by \citet{Tikhonov15} and later was
corrected by \citet{Anand19}.
Giving this information, we can consider in principle both options of the distance to Cas~I.
If Cas~I is situated at the distance close to that of IC342, its projected distance
to IC342  correponds to $\sim$0.29~Mpc. If, alternatively, it is situated closer to
Maffei1, the respective projected distance is $\sim$0.35~Mpc.

Thus, if we use only celestial coordinates and the radial velocity of Cas~I, based
on the probable membership in groups of either IC342, or of Maffei1/Maffei2,  we
expect its distance to be close either to 3.3 or to 5.7~Mpc. On the other hand,
there are two descripant direct distance determinations via the TRGB method:
D = 1.6$\pm$0.1~Mpc \citep{Tikhonov19} and 4.5$\pm$0.20~Mpc (EDD), which fall well
outside of the environs of these aggregates.
In Sections~\ref{ssec:O/H}, \ref{ssec:distance}, we use our estimate of gas O/H in Cas~I
and the improved estimates of the MW extinction in order to add arguments in favour
some of these variants.

\subsection{Gas metallicity versus the global parameters and the expected M$_{\rm B}$}
\label{ssec:O/H}

The gas metallicity of late-type galaxies in the LV follows the trend
described by the relation of 'O/H versus M$_{\rm B}$' from \citet{Berg12}.
See its illustration in Fig.~\ref{fig:ZvsL_Cas1}.
The respective linear regression reads as 12+log(O/H) = 6.272 -- 0.107 $\times$ M$_{\rm B}$,
with the rms scatter of log(O/H) of 0.15~dex. It extends over the range of
M$_{\rm B}$ = [--9.0,--19.0]. The great majority of this LV reference sample
belongs to typical groups and their close environs. As shown in
\citet{PaperVII,XMP-BTA},
the late-type dwarfs in the nearby voids have, on average, the reduced
values of log(O/H) by 0.14~dex (or by $\sim$30 percent, with the
rms scatter of 0.18~dex) relative to this reference relation. This
finding was interpreted as an evidence of the slower galaxy  evolution
in voids. Consistently with this idea, void galaxies have also
the elevated \HI\ content, on average by 40 percent \citep{PaperVI}.

As mentioned in Introduction,  Cas~I did not fall within the nearby voids defined in \citet{PTM19}.
However, if it would be situated at D = 1.6~Mpc, it should be a void object and one could expect its reduced metallicity.
% it appears a more isolated than for the case of its membership with IC342 or Maffei1.
In any case, we expect that the above relation of \citet{Berg12} is applicable
to Cas~I at first approximation. Thus, having its gas O/H with a good accuracy, we can get an estimate
of the most probable range of its M$_{\rm B}$.
Adopting the average metallicity of two \HII\ regions as 12+log(O/H) = 7.71$\pm$0.10~dex,
we derive the expected value of M$_{\rm B}$ = --13.44 $\pm$ 1.68~mag.
The latter uncertainty accounts for the r.m.s. scatter of the above \citet{Berg12}
relation (0.15~dex) and the accuracy of the derived average O/H in Cas~I.

\subsection{Implications for the distance of Cas~I}
\label{ssec:distance}

In the further estimate we use the most probable range of the absolute blue magnitude corresponding
to the gas metallicity of Cas~I from the previous section. To derive its distance modulus and the
respective distance, we adopt its total blue magnitude B$_{\rm tot}$ = 15.30,
as in Table~\ref{tab:parameters}. Then, the corrected for the Milky Way extinction, with the adopted
A$_{\rm B}$ = 3.48$\pm$0.16~mag, B$_{\rm tot,c}$ it equal to 11.82 $\pm$0.16~mag.
The respective distance modulus (m--M) = 11.82 + 13.44  = 25.26$\pm$1.68~mag.
This m--M corresponds to the distance derived from the relation of \citet{Berg12}, D(O/H) = 1.27~Mpc. The large
1$\sigma$ uncertainty in m--M corresponds to a large uncertainty of D(O/H) by a factor of $\sim$2.17.
That is the range $\pm$1$\sigma$=1.68~mag in m--M, corresponds to the range of D(O/H) from 0.58 to 2.75~Mpc.
Thus, despite the gas metallicity of Cas~I allows a very wide range of its probable distance, this can provide
the additional arguments for the choice between the various distance options.
% 0.461, 2.168

First, this metallicity-based distance, D(O/H), is too hard to agree with the distance of $\sim$5.7~Mpc
to the Maffei~1,2 group.
Second, we adopt that the recent EDD determination of the Tip of RGB in $I$-band, m$_{\rm I}$(TRGB) = 25.75,
due to its deepness, is much more reliable than that of \citet{Tikhonov19}.
We also adopt as a first approximation, the corrected MW extinction, derived in this work from the Balmer decrement
(A$_{\rm B}$ = 3.48$\pm$0.14). This implies the reduced extinction A$_{\rm I}$ = 1.447 in comparison to the EDD value
of 1.532.
This difference of 0.085~mag transforms to the reduction by 4.2 percent of D$_{\rm TRGB}$(EDD) =4.51~Mpc.
That is an improved  estimate of the MW extinction gives an independent estimate of D$_{\rm TRGB}$ = 4.32$\pm$0.39~Mpc.

However, the latter value of A$_{\rm B}$ = 3.48~mag should be treated as a strict upper limit, since it does not
account for the
internal extinction in the studied \HII\ regions. The internal extinction in extragalactic \HII\ regions is commonly not
large, typically of C(H$\beta$)$\lesssim$0.2 \citep[e.g.][]{Guseva17}. If we adopt for the two \HII\ regions
in Cas~I, a median internal C(H$\beta$)$\sim$0.1   \citep[as in][]{Guseva17},
we need to further reduce the measured above C(H$\beta$)=1.25 by 0.10, or by $\sim$8~per cent of this C(H$\beta$),
in order to treat this as a purely MW extinction.
This changes the adopted above value A$_{\rm I}$ = 1.447~mag, to the new one of 1.340~mag, that is by
0.107~mag smaller. This further reduces the D$_{\rm TRGB}$ = 4.32 by $\sim$5~per cent. Thus, the doubly corrected
original D$_{\rm TRGB}$(EDD)=4.51~Mpc  transforms to  4.11$\pm$0.36~Mpc.
This correction also applies to A$_{\rm B}$, resulting in the value of 3.20~mag. This reduces the corrected B$_{\rm tot,c}$
to 13.10$\pm$0.16 and increases the parameter m--M to 25.54~mag. So that the related distance D(O/H) shifts to 1.64~Mpc.

Summarizing, we arrive to two independent distance estimates of Cas~I, D$_{\rm TRGB}$ = 4.11$\pm$0.36~Mpc and
D(O/H) = 1.64~Mpc, with 1$\sigma$ uncertainty of factor of 2.17.
In the probabilistic approach, they can be consistent each to other, if the real value of D(Cas~I) is somewhere midway with
the account of the strongly different 1$\sigma$ error for the two methods. Namely, the values of
D(O/H) $\times$1~$\sigma$ = 3.56~Mpc, and D$_{\rm TRGB}$ -- 1~$\sigma$ = 3.75~Mpc are very close.
So that, at the first approximation, albeit with some tension, D(Cas~I) $\sim$ 3.65~Mpc will be the best match
to the both independent estimates of this parameter.
This naturally assigns Cas~I to the environs of IC342.

In Table~\ref{tab:parameters} we summarise the basic parameters of Cas~I as well as parameters, discussed in the paper,
that allows to catch them in one glimps.

\section[]{Summary and conclusions}
\label{sec:summary}

Summarising the  data, analysis, and discussion presented above, we arrive
at the following conclusions.

\begin{enumerate}
\item
We conducted BTA spectroscopy of two \HII\ regions in the LV dIrr galaxy Cas~I
($\#$2 and $\#$4 from \citet{Tikhonov19})
and measured their Oxygen abundance of 12+log(O/H) = 7.83 and 7.59$\pm$0.10~dex,
respectively. Their average of 12+log(O/H) = 7.71 correponds to the gas metallicity of Z(gas) $\sim$ Z\sunn/10,
a factor of 3--5 higher than the extremely low metallicity of stars as estimated by \citet{Tikhonov19}, Z(stars)
$\sim$0.02--0.03~Z\sunn.
\item
We use the observed Balmer decrement in the obtained spectra to address the issue of the large MW extinction.
The total extinction, derived via Balmer decrement, is qualitatively consistent but somewhat smaller and of the higher
accuracy than the large extinction derived via the dust IR emission by \citet{Schlegel98, Schlafly11}.
We then additionally account for a small internal extinction typical of \HII\ regions C(H$\beta$)(\HII) $\sim$0.10,
of A$_{\rm B}$(\HII) $\sim$0.28 mag. This finally results in A$_{\rm B}$(Balmer) = 3.06$\pm$0.14
versus A$_{\rm B}$(dust)=3.69$\pm$0.4, about 20~per~cent lower than adopted for the distance estimate in EDD.
This way improved estimate of the MW extinction is applied to correct the distance of Cas~I derived via the TRGB method.
This re-evaluation of the MW extinction results in the reduction of D(TRGB) to Cas~I from 4.51 to 4.11$\pm$0.36~Mpc.
\item
Gas metallicity in the LV late-type galaxies, according to \citet{Berg12}, is rather tightly related to
their luminosity and mass. We use the derived estimate of O/H in Cas~I to bracket its blue absolute magnitude using
the reference relation of 12+log(O/H) versus M$_{\rm B}$ from \citet{Berg12}. This gives us M$_{\rm B}$(Cas~I) =
--13.44$\pm$1.68~mag. Having the total apparent B$_{\rm tot}$, one obtains an independent estimate of the distance
modulus (m--M), and the related distance to Cas~I, D(O/H) $\sim$1.64~Mpc, with the 1$\sigma$ uncertainty of multiplication
factor 2.17.
\item
Both distance estimates, D(TRGB) = 4.11$\pm$0.2~Mpc and D(O/H) = 1.64~Mpc, can be consistent each to other with
a small tension (with $\sim$1~$\sigma$ downward and upward, respectively), at the average of D(Cas~I) $\sim$3.65~Mpc.
The latter distance estimate clearly favours the hypothesis on Cas~I to reside in the environs of IC342.
\end{enumerate}

\section*{Acknowledgements}
The reported study was funded by Russian Science Foundation according to the
research project  22-22-00654.
The authors thank N.A.~Tikhonov for sharing his results on very low metallicity
of stars in Cas~I  prior publication.
Observations with the SAO RAS telescopes are supported by the Ministry of Science and
Higher Education of the Russian Federation. The renovation of telescope equipment is
currently provided within the national project "Science and Universities".
This research is partly based on observations made with the NASA/ESA Hubble
Space Telescope obtained from the Space Telescope Science Institute,
which is operated by the Association of Universities for Research in
Astronomy, Inc., under NASA contract NAS 5-26555. These observations are associated
with program SNAP-15922.
We acknowledge the use for this work of the database HyperLEDA\footnote{http://leda.univ-lyon1.fr}.
This research has made use of the NASA/IPAC Extragalactic Database (NED)
which is operated by the Jet Propulsion Laboratory, California Institute
of Technology, under contract with the National Aeronautics and Space Administration.

\section*{Data availability}
%The data underlying this article are available
%in the The Extragalactic Distance Database (EDD)\footnote{http://edd.ifa.hawaii.edu/}.
The HST/ACS data used in this article are available in the
STScI data archive.

%===========================================================================

\label{lastpage}


\begin{thebibliography}{99}

\bibitem[\protect\citeauthoryear{Afanasiev \& Moiseev}{2005}]{SCORPIO}
  Afanasiev V.L., Moiseev A.V., 2005, Astron. Lett., 31, 193

%\bibitem[\protect\citeauthoryear{Afanasiev \& Moiseev}{2011}]{SCORPIO-2}
%  Afanasiev V.L., Moiseev A.V., 2011, BaltA, 20, 363

\bibitem[\protect\citeauthoryear{Anand et al.}{2019}]{Anand19}
  Anand G.S., Tully R.B., Rizzi L.,  Karachentsev I.D.,
  2019, ApJ Letters, 872, L4 % (15pp)
% The Distance and Motion of the Maffei Group


\bibitem[\protect\citeauthoryear{Anand et al.}{2021}]{Anand21}
  Anand G.S., Rizzi L., Tully R.B. et al.
  2021, AJ, 162, 80 % (15pp)
% The Extragalactic Distance Database: The Color-Magnitude
% Diagrams/Tip of the Red Giant Branch Distance Catalog
% Anand G.S., Rizzi L., Tully R.B., Shaya E.D., Karachentsev I.D.,
% Makarov D.I., Makarova L., Wu Po-Peng, Dolphin A.E, Kourkchi E.

%\bibitem[\protect\citeauthoryear{Anand et al.}{2021}]{Anand2021}
%  Anand G.S., Lee J.C., Van Dyk S.D., et al.
%  2021, MNRAS, 501, 3621 % (15pp)
% Distances to PHANGS galaxies: New tip of the red giant branch
% measurements and adopted distances.

\bibitem[\protect\citeauthoryear{Asplund et al.}{2009}]{Asplund2009}
Asplund M., Grevesse N., Sauval A.J. Scott P.,
   2009, Ann.Rev.Astron.Astrophys., 47, 481
% The Chemical Composition of the Sun

\bibitem[\protect\citeauthoryear{Berg et al.}{2012}]{Berg12}
    Berg D.A., Skillman E., Marble A. et al.
   2012, ApJ, 754, 98
  % Direct Oxygen Abundances for Low-luminosity LVL Galaxies

%\bibitem[\protect\citeauthoryear{Ekta, Chengalur \& Pustilnik}{Ekta et al.}{2008}]{Ekta08}
%  Ekta, Chengalur J.N., Pustilnik S.A., 2008, \mnras, 391, 881

\bibitem[\protect\citeauthoryear{Filippenko}{1982}]{Filippenko82}
Filippenko A.V., 1982, PASP, 94, 715

\bibitem[\protect\citeauthoryear{Guseva et al.}{2017}]{Guseva17}
   Guseva N.G., Izotov Y.I., Fricke K.J., Henkel C.,
   2017, A\&A, 599, A65
%  DR12 metal deficient candidates

%\bibitem[\protect\citeauthoryear{Haynes et al.}{2018}]{ALFALFA18}
%  Haynes M.P. et al., 2018, ApJ, 861, 49
% The Arecibo Legacy Fast ALFA Survey (final)

%\bibitem[\protect\citeauthoryear{Huchtmeier et al.}{2000}]{HKKE2000}
%    Huchtmeier W., Karachentsev I.D., Karachentseva V.E., Ehle M.,
%    2000, A\&A Suppl., 141, 469

\bibitem[\protect\citeauthoryear{Izotov et al.}{1994}]{Izotov94}
Izotov Y.I., Thuan T.X., Lipovetsky V.A., 1994,  \apj,  435, 647

%\bibitem[\protect\citeauthoryear{Izotov \& Thuan}{1999}]{Izotov99}
%   Izotov Y.I., Thuan T.X., 1999, ApJ,  639
%  Heavy-Element Abundances in Blue Compact Galaxies

\bibitem[\protect\citeauthoryear{Izotov et al.}{2006}]{Izotov06}
Izotov Y.I., Stasi\'{n}ska G., Meynet G., et al.,
2006, A\&A, 448, 955

\bibitem[\protect\citeauthoryear{Izotov \& Thuan}{2007}]{IT07}
Izotov Y.I., Thuan T.X., 2007,  \apj,  665, 1115


%\bibitem[\protect\citeauthoryear{Jacobs et al.}{2009}]{Jacobs09}
%   Jacobs B.A., Rizzi L., Tully R.B., Shaya E.J., Makarov D.I., Makarova L.N.,
%   2009, 138, 332
%  The Extragalactic Distance Database: Color-Magnitude Diagrams


\bibitem[\protect\citeauthoryear{Kaisin \& Karachentsev}{2019}]{Kaisin2019}
   Kaisin S.S., Karachentsev I.D., 2019, Astropys.Bull., 74, 1
% Star formation in nearby dwarf galaxies

%\bibitem[\protect\citeauthoryear{Karachentseva \& Karachentsev}{1998}]{KK1998}
%   Karachentseva V.E., Karachentsev I.D., 1998, A\&A Suppl., 127, 409

\bibitem[\protect\citeauthoryear{Karachentsev et al.}{1997}]{Karachentsev97}
     Karachentsev I., Drozdovsky I., Kaisin S., Takalo L.O., Heinamaki P., Valtonen M.,
     1997, A\&AS, 124, 559
% Revised photometric distances to nearby dwarf galaxies in the IC 342/Maffei complex

\bibitem[\protect\citeauthoryear{Karachentsev et al.}{2013}]{UNGC}
     Karachentsev I.D., Makarov D.I., Kaisina E.I., 2013, AJ, 145, 101
%  and its updates at
% \mbox{https://www.sao.ru/lv/lvgdb/object.php?name=Cas1\&id=54}
% Updated Nearby Galaxy Catalog

\bibitem[\protect\citeauthoryear{Pilyugin \& Thuan}{2005}]{PT05}
Pilyugin L.S., Thuan T.X., 2005, ApJ, 631, 231

%\bibitem[\protect\citeauthoryear{Pilyugin et al.}{2012}]{Counterpart2012}
%      Pilyugin L.S., Grebel E.K., Mattsson L., 2012, MNRAS, 424, 2316

\bibitem[\protect\citeauthoryear{Pustilnik, Martin}{2016}]{PaperVI}
      Pustilnik S.A., Martin J.-M., 2016, A\&A, 596, A86

%\bibitem[\protect\citeauthoryear{Pustilnik, Tepliakova}{2011}]{PaperI}
%      Pustilnik S.A., Tepliakova A.L., 2011, MNRAS, 415, 1188

\bibitem[\protect\citeauthoryear{Pustilnik, Perepelitsyna \& Kniazev}{Pustilnik et al.}{2016}]
  {PaperVII}
 Pustilnik S.A., Perepelitsyna Y.A., Kniazev A.Y., 2016, MNRAS, 463, 670

\bibitem[\protect\citeauthoryear{Pustilnik, Tepliakova \& Makarov}{Pustilnik et al.}{2019}]{PTM19}
   Pustilnik S.A., Tepliakova A.L., Makarov D.I., 2019, MNRAS, 482, 4329
% Void galaxies in the nearby Universe. I. Sample description

\bibitem[\protect\citeauthoryear{Pustilnik et al.}{2020}]{XMP-SALT}
Pustilnik S.A., Kniazev A.Y., Perepelitsyna Y.A., Egorova E.S.,
  2020, MNRAS, 493, 830

\bibitem[\protect\citeauthoryear{Pustilnik et al.}{2021}]{XMP-BTA}
Pustilnik S.A., Egorova E.S., Kniazev A.Y., Perepelitsyna Y.A., Tepliakova
   A.L., Burenkov A.N., Oparin D.V., 2021, MNRAS, 507, 944

\bibitem[\protect\citeauthoryear{Schlafly \& Finkbeiner}{2011}]{Schlafly11}
  Schlafly E.F., Finkbeiner D.P., 2011, ApJ, 737, 103 % 13pp.

\bibitem[\protect\citeauthoryear{Schlegel, Finkbeiner, Douglas}{Schlegel et al.}{1998}]{Schlegel98}
   Schlegel D.J., Finkbeiner D.P., Douglas M., 1998, ApJ, 500, 525

%\bibitem[\protect\citeauthoryear{Skillman}{1987}]{skillman1987} Skillman E.D.,
%  1987, in NASA Conf. Publ Vol. 2466, Star Formation in Galaxies, Washington, p.~263

\bibitem[\protect\citeauthoryear{Stasi\'nska}{1990}]{Stasinska90}
     Stasi\'nska G., 1990, A\&AS, 83, 501
% A grid of model HII regions for extragalactic studies


\bibitem[\protect\citeauthoryear{Tikhonov}{1996}]{Tikhonov96}
      Tikhonov N.A., 1996, Astronomische Nachrichten, 317, 175

\bibitem[\protect\citeauthoryear{Tikhonov, Galazutdinova}{2015}]{Tikhonov15}
      Tikhonov N.A., Galazutdinova O.A., 2015, Astropysical Bulletin, 73, 279
% Does the IC 342/Maffei Galaxy Group Really Exist?

\bibitem[\protect\citeauthoryear{Tikhonov}{2019}]{Tikhonov19}
      Tikhonov N.A., 2019, Astropysical Bulletin, 74, 396
%   Cas I

\bibitem[\protect\citeauthoryear{Weinberger}{1995}]{Weinberger95}
 Weinberger R., 1995, PASP, 107, 58
%  Discovery of Cas~I (not under this name)

\bibitem[\protect\citeauthoryear{Weinberger \& Suarer}{1998}]{WS1998}
Weinberger R., Suarer W., 1998, A\&A, 332, 523-525
% A bright emission region in the new nearby dI galaxy Cassiopea 1

\end{thebibliography}
\end{document}